\documentclass[
 reprint,
 amsmath,amssymb,
 prl,
superscriptaddress
]{revtex4-2}

\usepackage{graphicx}
\usepackage{dcolumn}
\usepackage{bm}
\usepackage{enumitem}
\usepackage{braket}
\usepackage{graphicx}
\usepackage{amsmath}
\usepackage{mathrsfs}
\usepackage{xcolor}
\usepackage{longtable}
\usepackage{amsfonts}
\usepackage{tikz}
\usepackage[colorlinks,linkcolor=blue,citecolor=blue,anchorcolor=blue]{hyperref}
\usepackage{CJK}

\usepackage{amsmath}
\usepackage[normalem]{ulem}

\begin{document}

\title{Any-to-any connected cavity-mediated architecture for quantum computing with trapped ions or Rydberg arrays}

\author{Joshua Ramette}
\affiliation{Department of Physics, MIT-Harvard Center for Ultracold Atoms and Research Laboratory of Electronics, Massachusetts Institute of Technology, Cambridge, Massachusetts 02139, USA}
\author{Josiah Sinclair}
\affiliation{Department of Physics, MIT-Harvard Center for Ultracold Atoms and Research Laboratory of Electronics, Massachusetts Institute of Technology, Cambridge, Massachusetts 02139, USA}
\author{Zachary Vendeiro}
\affiliation{Department of Physics, MIT-Harvard Center for Ultracold Atoms and Research Laboratory of Electronics, Massachusetts Institute of Technology, Cambridge, Massachusetts 02139, USA}
\author{Alyssa Rudelis}
\affiliation{Department of Physics, MIT-Harvard Center for Ultracold Atoms and Research Laboratory of Electronics, Massachusetts Institute of Technology, Cambridge, Massachusetts 02139, USA}
\author{Marko Cetina}
\affiliation{Duke Quantum Center and Department of Physics, Duke University, Durham, NC 27708, USA}
\author{Vladan Vuleti\'c}
 \affiliation{Department of Physics, MIT-Harvard Center for Ultracold Atoms and Research Laboratory of Electronics, Massachusetts Institute of Technology, Cambridge, Massachusetts 02139, USA}

\date{\today}

\begin{abstract}
We propose a hardware architecture and protocol for connecting many local quantum processors contained within an optical cavity. The scheme is compatible with trapped ions or Rydberg arrays, and realizes teleported gates between any two qubits by distributing entanglement via single-photon transfers through a cavity. Heralding enables high-fidelity entanglement even for a cavity of moderate quality. For processors composed of trapped ions in a linear chain, a single cavity with realistic parameters successfully transfers photons every few $\mu$s, enabling the any-to-any entanglement of 20 ion chains containing a total of 500 qubits in 200 $\mu$s, with both fidelities and rates limited only by local operations and ion readout. For processors composed of Rydberg atoms, our method fully connects a large array of thousands of neutral atoms.
The connectivity afforded by our architecture is extendable to tens of thousands of qubits using multiple overlapping cavities, expanding capabilities for NISQ era algorithms and Hamiltonian simulations, as well as enabling more robust high-dimensional error correcting schemes.
\end{abstract}

\maketitle

Qubit connectivity is an essential ingredient for extracting quantum advantage from a quantum computer. Whereas information propagation within a locally connected device of dimension $D$ is limited to an effective light cone, requiring a gate depth scaling as $N^{1/D}$ for scrambling across $N$ qubits, all-connected systems can scramble information with gate depth logarithmic in $N$ \cite{sekino2008, brown2013, boixo2018}. Thus, nonlocal gates enable faster exploration of the device's full Hilbert space than locally connected devices, whose computational power saturates when distant qubits cannot be entangled before decoherence sets in \cite{zhou2020}. Nonlocal gates would expand the capabilities of Noisy Intermediate Scale Quantum (NISQ) devices to implement algorithms and simulate Hamiltonians \cite{altman2021} with nonlocal terms ranging from qubit-based encodings of the Fermionic molecular Hamiltonian \cite{bauer2019} to dual quantum gravity models \cite{garcia2017} and other systems with qualitatively different dynamical behaviors due to fast information scrambling \cite{bravyi2006, foss2015, bentsen2019}. Beyond the NISQ era, full connectivity is favorable for quantum error correction, increasing error thresholds \cite{svore2005} and enabling high-dimensional codes with favorable scalings of logical qubit encoding rates and codeword distances
\cite{baspin2021, breuckmann2020, bravyi2013, Bravyi2009}.

\begin{figure}[b]
\includegraphics[width=8.6cm]{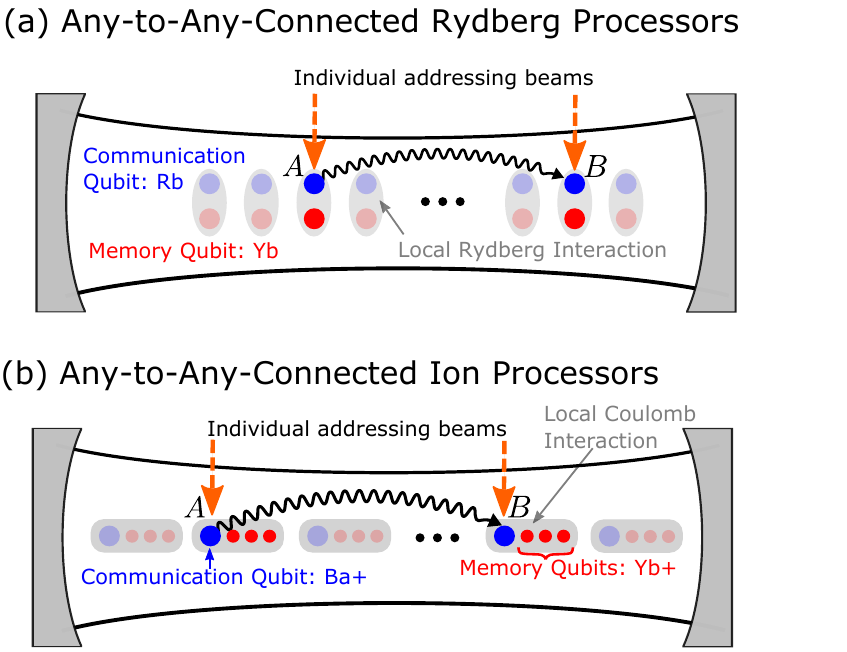}
\caption{Architecture for cavity-mediated quantum gate between any two qubits applied to (a) Rydberg atoms or (b) multiple chains of trapped ions. Photon transfers via the cavity produce heralded entanglement of communication qubits. Local interactions between communication and memory qubits enable teleported gates between any two memory qubits.  Different atomic species can be used for the communication and memory qubits so that the latter remain unaffected during the entanglement process for communication qubits.}
\label{cavity}
\end{figure}

Connectivity inherently conflicts with scalability, since having both requires either a high-dimensional topology or infinite-range interactions. For example, while in Rydberg atom arrays qubit numbers have rapidly increased to over 250 \cite{henriet2020, schymik2020, scholl2021, semeghini2021, bluvstein2021}, connectivity so far remains limited to a few neighboring atoms within the Rydberg blockade radius \cite{ Bernien2017, Levine2019, bluvstein2021}. In contrast, linear trapped ion chains, which natively support full connectivity via the collective motion of the chain, have seen slow increases in qubit number due to gate fidelities degrading with chain size \cite{cetina2020}, and challenges connecting multiple ion chains together. Current experiments have demonstrated intermodule connection based on ion shuttling \cite{ monroe2014, pino2021} and photonic interconnects \cite{monroe2013,Northup2014, hucul2015, Brown2016, inlek2017, stephenson2020} with rates far below local operation speeds (though faster rates might be reached with cavity-enhanced collection \cite{takahashi2017, takahashi2020,schupp2021}). As a result, only very small fully connected NISQ devices have been experimentally realized \cite{friis2018, cetina2020, pogorelov2021}.

In this Letter, we propose a new architecture for quantum computing which takes small, high-fidelity, local quantum processors, places them inside an optical cavity, and connects them using heralded single-photon transfers.
We examine specific implementations of our approach with a cavity of moderate volume and quality for both trapped ions and Rydberg arrays. First, we show that we can realize a fully-connected trapped ion system containing 20 chains and 500 ions with both gate fidelity and speed limited by local Coulomb operations and state readout, rather than the ion-cavity coupling. For Rydberg arrays, we show that our scheme can be used to provide arbitrary connectivity  for up to 1500 neutral-atom qubits.
Our architecture is extendable to multiple overlapping cavities containing tens of thousands of trapped ion or Rydberg qubits. It also enables state readout for individual qubits within a few ten of microseconds.

Previous proposals to engineer gates via optical cavities have been susceptible to photon losses, and have consequently required stringent experimental paramters in terms of cavity finesse and/or high-fidelity single-photon sources. Schemes enacting nonlocal gates through direct transfers of photons between atoms \cite{Pellizzari1995} have errors which scale poorly with the cooperativity ($C$) as $1/\sqrt{C}$ \cite{Sorensen2003}, or success probabilities with similar scaling for heralded schemes \cite{Borregaard2015}. Alternative schemes where the phase of a reflected photon flips depending on the internal state of atoms coupled to a cavity mode \cite{Duan2005, Sun2018, Wade2016} require high-fidelity single-photon sources. In contrast, our scheme uses a teleported gate protocol to avoid a direct coupling of the quantum register to the cavity, protecting the quantum information from the decoherence associated with optical losses.

\begin{figure}[h]
\includegraphics[width=8.6cm]{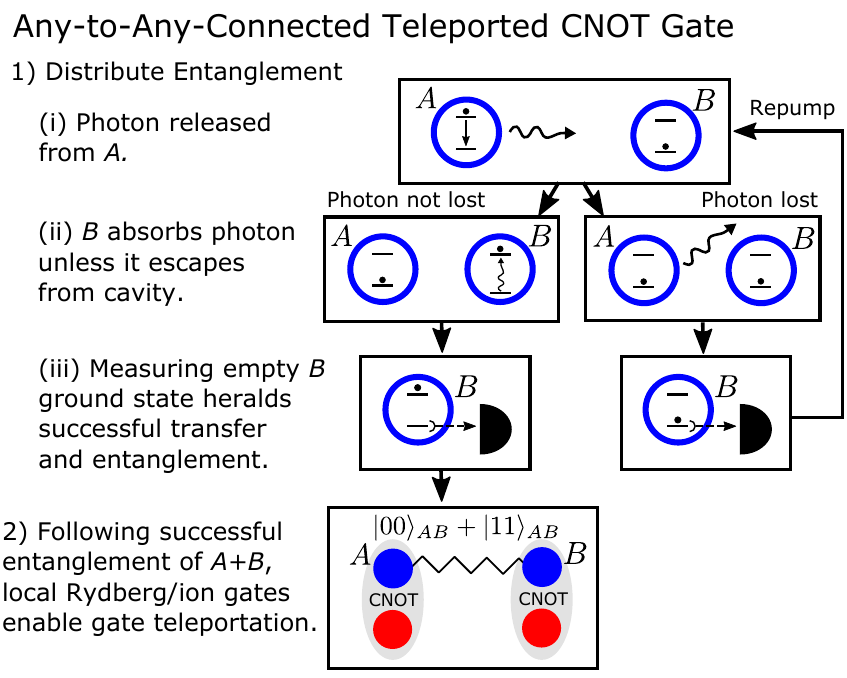}
\caption{Heralded photon transfers create entanglement between distant communication qubits. Subsequent high-fidelity local operations enable teleported quantum gates between distant memory qubits.}
\label{herald}
\end{figure}

In Fig. \ref{cavity} (a) and (b), we consider a set of local processors composed of communication qubits and memory qubits. Communication qubits are coupled to the cavity and used for distribution of entanglement, while memory qubits couple to communication qubits via a local interaction (Rydberg interaction shown in (a) and Coulomb interaction shown in (b)). Photon loss processes associated with the cavity linewidth $\kappa$ and atomic state scattering associated with the atomic linewidth $\Gamma$ compete with the atom-cavity coupling $g$ to determine the success probability for entangling communication qubits. Each local processor is referred to as a node, such that in the figure, nodes $A$ and $B$ represent any two local processors within the cavity mode volume. Executing a teleported CNOT gate \cite{Gottesman1999, Chou2018} between any two qubits within our architecture requires the following steps, shown in Fig. \ref{herald}:

\begin{enumerate}[label=\arabic*)]
   \item Create a Bell state between the communication qubits at nodes $A$ and $B$ by (i)-(ii) passing a photon from $A$ to $B$ via a (lossy) cavity. With a carefully designed transfer protocol, measuring an empty ground state of $B$ ensures, without destroying the established entanglement, that the photon was not lost due to cavity or atomic excited-state decays. This leaves the communication qubits at $A$ and $B$ projected into a Bell state (iii).
   \item Once the non-local Bell pair has been established, enact local CNOT gates between communication and memory qubits at nodes $A$ and $B$, then measure the states of the $A$ and $B$ communication qubits. Based on the results, enact single qubit gates to realize a teleported CNOT gate between memory qubits at nodes $A$ and $B$.
\end{enumerate}

\section{\label{sec:level1} Entanglement Distribution Scheme}
\begin{figure}[b]
\includegraphics[width=8.6cm]{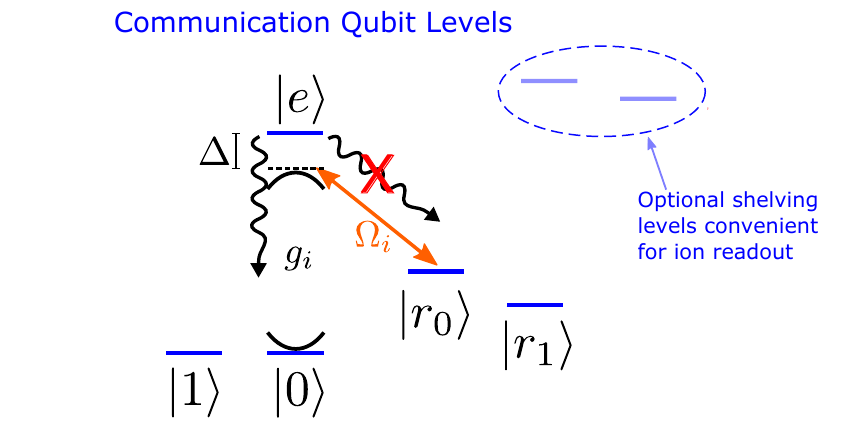}
\caption{Atomic level scheme for the communication qubits. The communication qubits are coupled to the cavity with strength $g_i$ and addressed individually with Rabi frequency $\Omega_i$. The excited state $|e \rangle$ decays at $\Gamma$, and the level and addressing schemes are chosen to suppress the probability of decay into the operation states $|r_0 \rangle$ and $|r_1 \rangle$. Raman transitions or microwaves enable site selective rotations between the qubit states $|0 \rangle$ and $|1 \rangle$ and between the operation states $|r_0 \rangle$ and $|r_1 \rangle$. For trapped ions, long-lived shelving levels can be used as part of the heralding process. For Rydberg atoms, Rydberg levels conveniently act as operation states.}
\label{gate_levels}
\end{figure}

We now describe a protocol to establish a Bell pair between any two communication qubits at nodes $A$ and $B$ with, in principle, unit fidelity despite cavity ($\kappa$) and atomic-scattering ($\Gamma$) loss mechanisms. The level structure we assume for network and qubit atoms is shown in Fig. \ref{gate_levels}. We assume that while $|e \rangle$ has a linewidth $\Gamma$, scattering back into $|r_0 \rangle$ or $|r_1 \rangle$ (which we call ``operation states" and could be Rydberg states in a realization with neutral atoms) is strongly suppressed, so that all scattering and photon loss events are detectable by leaving $B$ in the ground states $|0 \rangle$ or $|1 \rangle$.

Our protocol to generate the Bell pair begins in the state
\begin{equation}
|\Psi\rangle = \frac{1}{\sqrt{2}}(|r_0 \rangle_A + |r_1 \rangle_A) |0 \rangle_B |\textrm{vac} \rangle_E
\end{equation}
where all other network atoms are stored in state $|1 \rangle$ so they do not couple to the cavity, and the environment $E$ of modes external to the cavity is in the vacuum state.
Then, a photon is transferred through the cavity from $A$ to $B$, conditioned on $A$ being in $|r_0 \rangle_A$. The transfer is done either using two-photon Raman $\pi$ pulses with nonzero detuning $\Delta$ to release a photon from $A$ and absorb it at $B$, or with resonant STIRAP by ramping up $\Omega_A$ while ramping down $\Omega_B$ \cite{Pellizzari1995}. During this transfer, the photon may leak out of the cavity, or may scatter off of $|e \rangle_A$ or $|e \rangle_B$. Therefore we have $|r_0 \rangle_A |0 \rangle_B |\textrm{vac}\rangle_E \rightarrow \alpha |0 \rangle_A |r_0 \rangle_B |\textrm{vac}\rangle_E + \beta |\textrm{Loss}\rangle $, where $|\textrm{Loss} \rangle$ represents a generic state of the form $\sum _{k, l, \mu}\alpha_{k l \mu}|k \rangle_{A} |l \rangle_{B} \hat{a}^{\dag}_{\mu} |\textrm{vac}\rangle_E $ where the photon has scattered into a mode $\mu$ outside the cavity and the atoms $A$ and $B$ are left in some collection of ground states indexed by $k$ and $l$. Since the ground states are only coupled via the cavity mode for the remainder of the protocol, once in the state $|\textrm{Loss} \rangle$, atoms $A$ and $B$ cannot evolve out of their ground states. Thus a state $|\textrm{Loss} \rangle$ maps to $|\textrm{Loss} \rangle$ under further operations in the protocol. The lossy transfer results in the state:

\begin{multline}
|\Psi\rangle = \frac{1}{\sqrt{2}}(\alpha |0 \rangle_A |r_0 \rangle_B |\textrm{vac} \rangle_E + \beta|\textrm{Loss} \rangle ) \\
+ \frac{1}{\sqrt{2}}|r_1 \rangle_A |0 \rangle_B |\textrm{vac} \rangle_E
\end{multline}

We repeat the transfer on the other component in the Bell state to symmetrize the losses. First swap the two states $|r_0 \rangle$ and $|r_1 \rangle$ on $A$ and $B$ and also swap the two ground states on $A$:
\begin{multline}\label{trans}
|\Psi\rangle = \frac{1}{\sqrt{2}}(\alpha |1 \rangle_A |r_1 \rangle_B |\textrm{vac} \rangle_E + \beta|\textrm{Loss} \rangle ) \\
+ \frac{1}{\sqrt{2}}|r_0 \rangle_A |0 \rangle_B |\textrm{vac} \rangle_E
\end{multline}
Then, execute a second identical STIRAP as before which affects only the last term in Eq. \ref{trans}. Assuming the loss is symmetrized by using the same temporal profiles for $\Omega_A$ and $\Omega_B$ as before, this results in the same values for $\alpha$ and $\beta$ (where $|\textrm{Loss}'\rangle$ denotes the state where the photon was scattered during the second STIRAP):
\begin{align}\label{bell}
|\Psi\rangle = & \frac{1}{\sqrt{2}}(\alpha |1 \rangle_A |r_1 \rangle_B |\textrm{vac} \rangle_E + \beta|\textrm{Loss} \rangle) \nonumber \\
& + \frac{1}{\sqrt{2}}(\alpha |0 \rangle_A |r_0 \rangle_B |\textrm{vac} \rangle_E + \beta |\textrm{Loss}' \rangle) \nonumber \\
= & \alpha \frac{1}{\sqrt{2}} (|1 \rangle_A |r_1 \rangle_B  + |0 \rangle_A |r_0 \rangle_B) |\textrm{vac} \rangle_E \nonumber \\
& + \beta \frac{1}{\sqrt{2}} (|\textrm{Loss} \rangle + |\textrm{Loss}'\rangle)
\end{align}

Within the $\alpha$ component of $|\Psi\rangle$, $A$ and $B$ are now entangled.
As shown in Fig. \ref{herald}, measuring whether or not atom $B$ occupies a ground state projects $|\Psi\rangle$ either into $|\textrm{Loss} \rangle$ or into the Bell state $(|1 \rangle_A |r_1 \rangle_B  + |0 \rangle_A |r_0 \rangle_B)/\sqrt{2}$.

\section{\label{sec:exp}Experimental Implementations}

We demonstrate the feasibility of our approach by outlining specific implementations using commonly trapped species of ions and Rydberg atoms. We simulate the transfer dynamics by coupling two atoms/ions to the cavity mode and addressing them via a STIRAP pulse sequence of duration $T$ with Gaussian pulses of peak strength $\Omega_0$. The success probability forms a landscape over the space of $\Omega_0$ and $T$ values, where each point in the space represents a trade-off between excited-state losses ($\Gamma$) and cavity losses ($\kappa$). With cavities comparable to those used in existing experiments, we show how to realize few-$\mu$s transfer times with reasonable success probability (40\%) even with cavity mode volumes large enough to trap many atoms/ions. We also describe the leading expected error sources of the nonlocal gates and how our scheme's symmetry improves the fidelity of the Bell state.

The scheme is designed to be robust against slow variations in the laser and cavity coupling strengths (which may slightly affect the success probability but not the heralded fidelity), and does not require identical coupling strengths at any of the nodes. This allows entanglement even between communication qubits trapped over a broad range of positions within a spatially varying cavity mode profile.

Broadly, the errors of the two local gates ($\epsilon_{local}$) as well as those of the two communication qubit readout measurements ($\epsilon_M$) and the infidelity of the distributed Bell state ($\epsilon_{Bell}$) all contribute to the overall nonlocal gate error: $\epsilon_\textrm{CNOT} = 2 \epsilon_{local} + 2 \epsilon_M + \epsilon_{Bell}$. However, our scheme's two repeated transfers through the same cavity mode symmetrize the amplitudes on the two components of the heralded Bell pair despite photon losses. This makes the process insensitive to asymmetries and slow drifts in the coupling strengths and requires repeatability of the couplings only on the few-$\mu$s timescale. While laser pulse strength and phase fluctuations could introduce imperfections, excellent phase coherence over the few microseconds of optical control used in our scheme is possible, as evidenced by the long optical qubit coherence times routinely demonstrated \cite{bruzewicz2019}. Our scheme can also be made first-order insensitive to laser amplitude fluctuations by choosing the optimal value for $\Omega_0$ within the success probability landscape (see Supplement).

\subsection{\label{sec:ion}Application to Trapped Ions}
An implementation for local trapped-ion processors could use Ba$^+$ ions as communication qubits and Yb$^+$ ions as memory qubits (these two species are favorable for implementing interspecies local operations \cite{inlek2017}, although many other species combinations are possible, such as using different Ba$^+$ isotopes for communication and memory). This does not introduce extra complexity, as a second ion species is likely already necessary for sympathetic cooling in any trapped ion system and can also function as our communication qubits \cite{inlek2017}. To realize the level scheme depicted in Fig. \ref{gate_levels}, we use the ${}^{137}$Ba$^+$ isotope with nuclear spin $I = 3/2$. The operation states $|0 \rangle$ and $|1 \rangle$ in Fig. \ref{gate_levels} correspond to the clock states $|F = 2, m_F = 0 \rangle$ and $|F = 1, m_F = 0 \rangle$ in the $6S_{1/2}$ ground-state manifold, and the states $|r_0 \rangle$ and $|r_1 \rangle$ correspond to $|F = 2, m_F = 0 \rangle$ and $|F = 0, m_F = 0 \rangle$ in the metastable $5D_{3/2}$ state manifold, while the excited state $|e \rangle$ is the state $|F = 3, m_F = 0 \rangle$ in $6P_{3/2}$. Then rotations between $|0 \rangle$ and $|1 \rangle$ and between $|r_0 \rangle$ and $|r_1 \rangle$ can be driven via Raman transitions \cite{crocker2019}.
The states $|r_0 \rangle$ and $|r_1 \rangle$ in $5D_{3/2}$ are chosen because of the small branching ratio from $6P_{3/2}$ to $5D_{3/2}$ \cite{kurz2008}, which strongly suppresses spontaneous decay back into $|r_0 \rangle$ and $|r_1 \rangle$.

The cavity is tuned on resonance with the $|0 \rangle \rightarrow |e \rangle$ transition at 455 nm, at which wavelength we would expect losses in the mirrors of $2 \times 10^{-5}$ to be achievable \cite{gangloff2015}.
A cavity of length 2.8 mm with a 13 $\mu$m Gaussian mode waist would have a Rayleigh range of $1170$ $\mu$m, a linewidth of $\kappa = 2 \pi × 340$ kHz, and would give a coupling strength between the ion and the cavity at the center of the mode of $g = 2 \pi × 5.8$ MHz. With an excited state linewidth of $\Gamma = 2 \pi × 25$ MHz, this gives a cavity cooperativity $C = g^2/\kappa \Gamma = 4.0$. While success probability asymptotically scales like $1/\sqrt{C}$ for large cooperativity, with a cooperativity on the order of unity, the coupling strength $g$ to the cavity mode is of similar strength to the geometric mean of the loss processes $\sqrt{\kappa \Gamma}$, meaning we expect on the order of half the transfer attempts to be successful. A simulated photon transfer with these parameters gives an efficiency of around $p = 0.40$ over 1.0 $\mu$s, making it possible to establish entanglement between any two Ba$^+$ ions on the microsecond timescale with a laser pulse profile with a peak strength of $\Omega_0 = 2 \pi \times 18$ MHz. A cavity with these parameters could be integrated with a surface ion trap oriented along the cavity axis \cite{Cetina2013}, with the ions located about $70 \: \mu$m above the surface, and with $200$ $\mu$m space between each mirror and the closest ion \cite{takahashi2017}. For chains of 25 ions, each of length 88 $\mu$m with 3.5 $\mu$m inter-ion spacing and $30 \: \mu$m of space between chains, this configuration would fit 20 chains along the cavity axis (for a total of 500 qubits) within the Rayleigh range of the cavity mode.

\begin{figure}
\includegraphics[width=8.6cm]{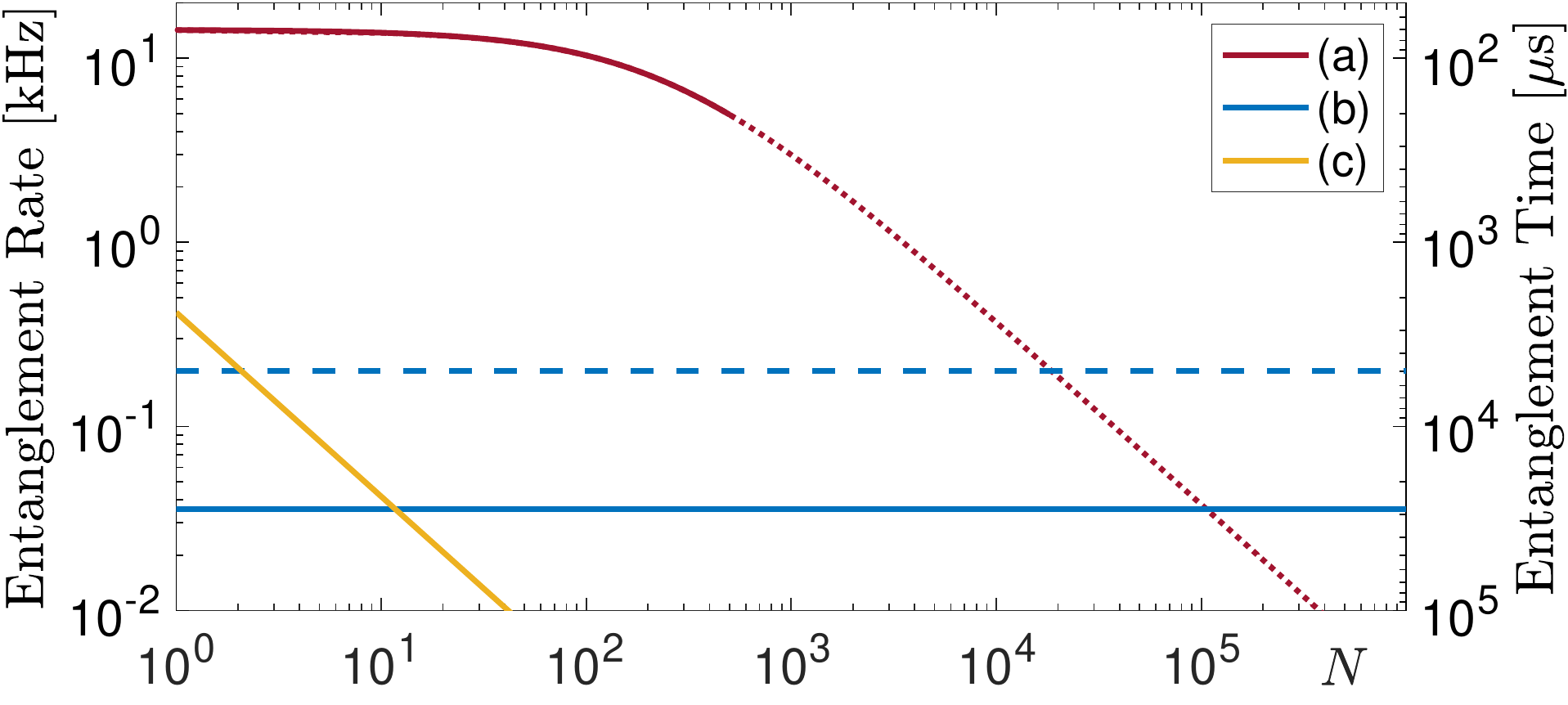}\vspace{3 mm}
\includegraphics[width=8.6cm]{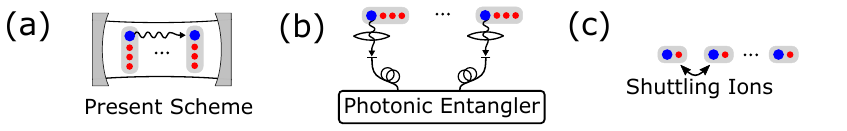}
\caption{Comparison of the average rates for distributing entanglement to $M$ any-to-any-connected local processors comprising $N$ total qubits for different modalities (a)-(c). (a) Red line: intracavity entanglement method proposed here with 25 ions per chain using the parameters outlined in this section, with the entanglement time given by $(\tau_\textrm{parallel} + \frac{1}{2} M \tau_\textrm{serial})/p$, and where $p = 0.40$, $\tau_\textrm{serial} = 5.3 \: \mu$s (see Supplement), and $\tau_\textrm{parallel} = 28 \: \mu$s \cite{noek2013}. The red line changes from solid to dashed at $N = 500$, representing the number of ions that could be fit comfortably in a single linear trap along the cavity axis. (b) Blue lines: photonic interconnects with 25 ions per chain using the method of Refs. \cite{inlek2017, olmschenk2009, hucul2015} can entangle many pairs of ion chains in parallel if the photons are routed through an optical cross-connect switch. Dashed blue shows the current state of the art for entangling two ions (5 ms) \cite{stephenson2020}, and solid blue shows the same with additional realistic overheads necessary for achieving high fidelity all-connectivity, including a factor of two overhead for entanglement purification \cite{stephenson2020, nigmatullin2016}, cross-connect switch insertion losses (1.3 dB per photon), and switching time (10 ms overhead) \cite{aksyuk2003, olkhovets2004}. (c) Yellow line: ion shuttling architectures \cite{Kielpinski2002}, where ions are physically transported between different ion traps, require milliseconds of sympathetic re-cooling after each split/merge operation. Demonstrated geometries (two ions per local processor, see \cite{pino2021}) lead to linearly increasing transport time for fully connected circuits.}
\label{ion_time_ions}
\end{figure}

Across all chains, the entanglement generation process would consist of a serial stage, where fast cavity transfers are carried out between all $M$ chains, followed by a stage of parallelized fluorescence readout to verify transfer success and herald entanglement.
Then, local operations in successfully heralded pairs, perfomed between communication and memory qubits, can directly use the entanglement for teleported gates, or can swap the entanglement onto other ions in the chain for later use. A full quantum gate circuit is implemented by repeated iterations of the above steps.

Following each fast cavity transfer in the serial stage, the logical states of both $A$ and $B$ can be shelved in $|5D_{5/2},F = 1, m_F = 0 \rangle$ and $|5D_{5/2},F = 2, m_F = 0 \rangle$ using Raman transitions. Shelving ensures atom $A$ does not couple to the cavity during subsequent transfers and prevents atom $B$ from scattering into the operation states during standard fluorescence readout on $6P_{1/2}$.
The entire attempt, including shelving, requires approximately $\tau_\textrm{serial} \approx 5$ $\mu$s  per pair of communication qubits (see Supplement).
Readout for heralding can then be done in parallel across all the pairs in a time $\tau_\textrm{parallel} \approx 28 \: \mu$s \cite{noek2013}.
On average, $pM$ of the chains are entangled in time $\tau_\textrm{parallel} + M \tau_\textrm{serial}/2$, so that the average amount of time taken to distribute $M/2$ Bell pairs among the $M/2$ pairs of chains is $(\tau_\textrm{parallel} + M \tau_\textrm{serial}/2)/p$.

For small $\tau_\textrm{serial}$, even for moderately large $M$ the entanglement distribution and nonlocal gate speed is on the order of local Coulomb gates and readout (tens of $\mu$s \cite{bruzewicz2019}).
Fig. \ref{ion_time_ions} shows how for the parameters described here, our scheme is limited primarily by the readout time $\tau_\textrm{parallel}$ for system sizes up to a few hundred qubits, and can distribute entanglement between 20 chains comprising 500 qubits in an average of 200 $\mu$s. The local gate errors are expected to dominate the overall teleported gate infidelity (see Supplement for discussion of mitigating contributions from entanglement preparation such as backscattering ($10^{-3}$), single-qubit rotations ($4 \times 10^{-3}$), and thermal motion ($10^{-2}$)) for the parameters outlined here. With the ability to link multiple chains at speeds and fidelities comparable to local gates, large trapped-ion quantum computers will enable NISQ applications far exceeding what is currently possible.

\subsection{\label{sec:Rydnerg}Application to Rydberg Arrays}

\begin{figure}
\includegraphics[width=8.6cm]{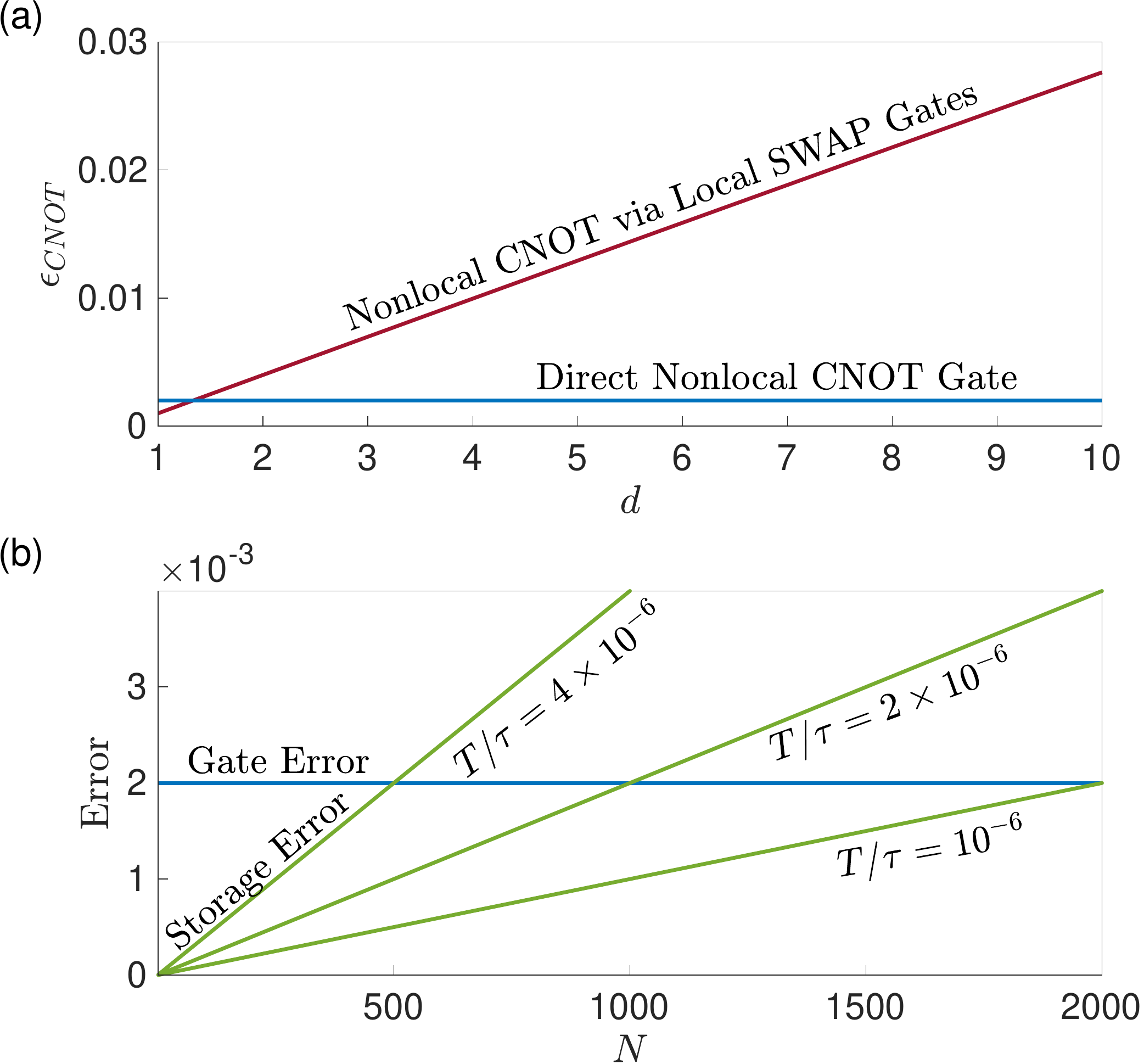}
\caption{With nearest-neighbor Rydberg gate errors of $\epsilon_R = 10^{-3}$ \cite{saffman_review_2016}, the nonlocal teleported CNOT gate error is limited to $\epsilon_{CNOT} \sim 2 \epsilon_R = 2 \times 10^{-3}$ (a) Error probabilities due to local two-qubit gate errors for gates between qubits $d$ sites apart, when implemented directly via teleportation (constant error $\epsilon_{CNOT}$, solid blue) or via local swap gates (red, 3 local CNOT gates per swap gate for an error $\epsilon_R + 3(d-1)\epsilon_R$). (b) Comparing storage and gate errors for implementing serial any-to-any gates. The serial system can be scaled until $N T /\tau \sim \epsilon_{CNOT}$ before storage errors would dominate gate errors.}
\label{rydberg_gates}
\end{figure}

As an example of a realization with realistic parameters in neutral atoms with local Rydberg gates, consider a dual-species experiment where for the communication qubits we use ${}^{87}$Rb (which can conveniently couple to a high-finesse cavity in the near infrared), and for the memory qubits we use ${}^{171}$Yb (where the magnetically insensitive, metastable nuclear spin states $|{}^{3}P_0, m_F = \pm 1/2 \rangle$ with lifetime $\tau = 26$ s function as logical states from which the Rydberg states are accessible via a single-photon transition). For the network ${}^{87}$Rb atoms, $|5S_{1/2}, F = 1, m_F = 0 \rangle$ and $|5S_{1/2}, F = 2, m_F =0 \rangle$ are the logical $|0 \rangle$ and $|1 \rangle$ states, and the cavity at 780 nm couples $|5S_{1/2}, F = 2, m_F =0 \rangle$ to $|e \rangle = $ $|5P_{3/2}, F=3, m_F=0 \rangle$. High-lying Rydberg states of the form $n S$ and $n P$ with $n\geq 70$ are used for $|r_0\rangle$ and $|r_1\rangle$, which are coupled by microwave radiation. We choose a cavity with a length $L = 2$ cm (where the centimeter scale cavity keeps the atoms far from the mirrors to mitigate any effects of stray electric fields \cite{georgakopoulos2018}) and moderate finesse $F = 140,000$ for a ringdown time $3.0 \: \mu$s and cavity linewidth $\kappa = 2 \pi \times$ 53 kHz. We assume a Gaussian mode with waist $10$ $\mu$m and coupling strength $g = 2 \pi × 2.8$ MHz, so that with the ${}^{87}$Rb natural linewidth of $2 \pi × 6$ MHz we have a cavity of moderate cooperativity $C = 25$ with a Rayleigh range of $z_R = 403$ $\mu$m. Simulating the STIRAP transfer efficiency with these cavity parameters, we achieve a successful transfer probability of over $p = 0.7$ in 2 $\mu$s with $\Omega_0 = 2 \pi × 8.6$ MHz. The total time spent to establish a Bell pair (two separate cavity-mediated transfers in the protocol) is then around $(1/0.7)\times2\times2$ $\mu$s $\approx 6$ $\mu$s.

The Rydberg states, already necessary for the local Rydberg gates, naturally realize the scattering suppression indicated in Fig. \ref{gate_levels}. The Rydberg states with typical lifetimes of a few hundred $\mu$s would occasionally decay during the few-$\mu$s transfer, but these small losses again can be symmetrized away and heralded with our scheme. This ensures that no distortion of the heralded Bell state occurs, as both components of the Bell state can be made to spend the same amount of time in the Rydberg states. Entanglement distribution failures are measured at atom $B$ following the transfer attempt by quickly depumping $F = 1$, then moving $|r_0\rangle_B$ and $|r_1\rangle_B$ into the $F = 1$ manifold and using a broadband laser and a single microwave frequency to couple the broad set of Rydberg states thermally accessible from $|r_0\rangle_B$ and $|r_1\rangle_B$ to the $F = 3$ excited state, from where it would decay to the $F = 2$ ground state. Transfer failure due to cavity losses or Rydberg decays then results in an $F = 2$ ground state atom $B$.

Non-destructive fluorescence imaging of dipole trapped neutral atoms typically takes milliseconds \cite{auger2017}, much slower than for ions.
However, the cavity itself can enable fast serial state readout \cite{gehr2010, Volz2011, boozer2006, mabuchi1996}. For example, probing the cavity transmission on the nearly closed $|5S_{1/2}, F = 2 \rangle$ to the $|5P_{3/2}, F=3 \rangle$ transition can be much faster, with readout speed only limited by the cavity ringdown time (here $3$ $\mu$s) and the necessity to avoid saturation \cite{gehr2010}  (which would only occur here with an incident probe photon rate beyond $~g^2/\kappa = 10^{3}$ $\mu$s$^{-1}$).

With a 2.6 $\mu$m spacing between sites \cite{endres2016} and assuming probabilistic loading of both species with no rearrangements and 50\% atom filling, such that the probability of loading both a ${}^{87}$Rb and a ${}^{171}$Yb is 25\%, this allows $\sim 75$ qubits in a 1D chain along the cavity mode axis within $\pm z_R$ (With rearrangement of both species this number would be 300 qubits.). One more row of qubits to each side of the axis could easily fit in a rectangular lattice within the 6 $\mu$m before the mode function falls of by $1/\sqrt{2}$ radially for a total of $\sim 225/900$ qubits without/with rearrangement. With rows in a triangular pattern \cite{Wang2020} with 2.6 $\mu$m lattice spacing, we could fit 5 total rows of atoms with no atoms more than 3.7 $\mu$m radially distant from the mode axis, for a total of $\sim 325/1500$ qubits without/with rearrangement.

In Fig. \ref{rydberg_gates} (a) we compare a Rydberg array limited to nearest-neighbor gates to our cavity-mediated architecture that represents a serial, any-to-any connected system.
The average nonlocal gate fidelity in the nearest-neighbor architecture degrades with increasing system size due to the increasing number of local SWAP gate errors.
In Fig. \ref{rydberg_gates} (b), we consider the effect of the nonlocal gate speed on qubit storage errors in a serial, any-to-any connected system. During each two qubit gate of duration $T$ with qubits of lifetime $\tau$, there is probability $N T /\tau$ of a qubit storage error among the $N$ qubits. 

Note that in locally connected systems, the SWAP gate overheads can lead not only to more local gate errors, but also more storage errors due to the extra time required to sequentially implement the SWAP gates.
In general, the time required to perform $N$ nonlocal gates across the $N$ qubits depends on both the degree of connectivity and parallelizability. Local systems in finite dimensions require increasing amounts of parallelization to equal or exceed the runtime and storage error performance of serial any-to-any connected systems (see Supplement for a more quantitative comparison).

\section{\label{sec:Conclusion}Conclusions and outlook}
\begin{figure}
\includegraphics[width=8.6cm]{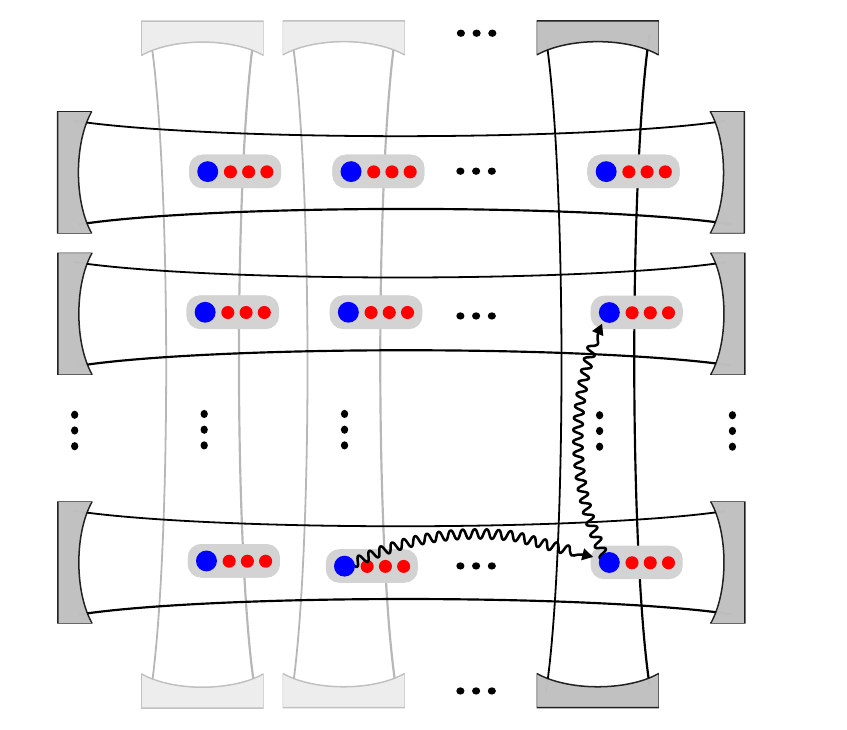}
\caption{Extension to larger systems. The methods outlined here could be extended using several overlapping cavity modes, creating systems with potentially $10^4$ all-connected qubits. Several vertical cavity modes (shaded) provide parallelization, such that many photon transfers can be accomplished simultaneously. Any two communication ions can be entangled in two steps, by photon-mediated communication along a horizontal and a vertical cavity mode.}
\label{2Darray}
\end{figure}

We have presented a new architecture for connecting many local quantum processors within a single optical cavity, and outlined feasible implementations of our scheme with local processors composed of linear chains of trapped ions or neutral Rydberg atoms. For both Rydberg atoms and trapped ions, our scheme makes it possible to create all-connected quantum systems of several hundred qubits within a single optical cavity with realistic parameters.

Our methods can easily be extended to multiple overlapping cavity modes with modest overheads in entanglement distribution speeds, allowing thousands of fully connected trapped ion or Rydberg qubits (see Fig. \ref{2Darray}).
With hundreds to thousands of fully connected qubits, many more NISQ simulations become feasible.
Beyond the NISQ era, any-to-any connectivity in systems with many qubits may allow the realization of logical qubits using some of the most promising quantum codes, and we hope that our proposal inspires additional theoretical investigation into optimal codes unlimited by connectivity constraints.
Once localized modules containing logical qubits are constructed, slower entanglement methods like two-photon entanglement distribution \cite{monroe2014, inlek2017, hucul2015} can then be used to connect those longer-lived logical-qubit modules into even larger systems.

\bibliography{ramette_biblio}

\end{document}


\preprint{APS/123-QED}

\title{Supplementary Material: Any-to-any connected cavity-mediated architecture for quantum computing with trapped ions or Rydberg arrays}

\author{Joshua Ramette}
\author{Josiah Sinclair}
\author{Zachary Vendeiro}
\author{Alyssa Rudelis}
\author{Marko Cetina}
\author{Vladan Vuletic}

\date{\today}

\maketitle

Here we give additional details regarding the possible implementations of our proposal for trapped ions described in the main text. First, we discuss a pulse sequence for generating entanglement on the timescale of a few microseconds, and then discuss the impact of experimental imperfections on the achievable entanglement fidelity.

\section{\label{sec:level1}Trapped Ion Implementation: Entanglement Speed}

To show that few-$\mu$s serial connection times $\tau_{serial}$ are realistic using current technology, we outline the details of a pulse sequence with ${}^{137}$Ba$^+$ ions trapped in a Paul trap, and budget the time taken for the pulses and transfers. As all the states used for transferring and shelving in the $5D_{5/2}$ manifold have magnetic quantum number $m_F = 0$, we can neglect all other Zeeman states by using $\pi$ polarized beams (the probability of populating other Zeeman states of $m_f \neq 0$ is $< 10^{-3}$ for a drive strength of $2\pi \times 3$ MHz, assuming $99\%$ polarization purity and $2\pi \times 10$ MHz Zeeman splitting). All transitions used are then between states of different $F$ quantum number, where off-resonant transitions are at least $2\pi \times 60$ MHz detuned. With $2\pi \times 3$ MHz Raman Rabi frequency, high fidelity Raman transitions of SK1 composite pulse sequences requiring $5\pi$ radians of rotation \cite{shappert2013, brown2004} can occur in $5\pi/(2\pi\times3$ MHz) = 0.83 $ \mu$s.

After parallelized state initialization, an entanglement attempt includes two sequential single-photon transfers (each 1 $\mu$s) and a single qubit rotation (0.83 $\mu$s). This pulse sequence entangles any two communication qubits $A$ and $B$, where $A$ is then in the $6S_{1/2}$ manifold and $B$ is in the $5D_{3/2}$ manifold. We can then quickly optically pump any population out of the $5D_{5/2}$ manifold ($ \ll 1 \: \mu$s) back to $6S_{1/2}$, in case it was populated by scattering during the transfers.
Following that, the qubit states of $A$ and $B$ are coherently shelved in the $|F = 2, m_F = 0 \rangle$ and $|F = 1, m_F = 0 \rangle$ states of the $5D_{5/2}$ manifold, respectively.
Qubit $A$ can be shelved using two Raman tones directly coupling $|0 \rangle_A$ and $|1 \rangle_A$ to $|5D_{5/2}, F = 2, m_F = 0 \rangle_A$ and $|5D_{5/2}, F = 1, m_F = 0 \rangle_A$ respectively.
Shelving qubit $B$ takes a few additional steps because of selection rules: first move $|r_0 \rangle_B$ $\rightarrow$ $|5D_{5/2}, F = 2, m_F = 0 \rangle_B$, then in parallel move $|r_1 \rangle_B$ $\rightarrow$ $|r_0 \rangle_B$ and $|5D_{5/2}, F = 2, m_F = 0 \rangle_B$ $\rightarrow$ $|5D_{5/2}, F = 1, m_F = 0 \rangle_B$, and finally now move $|r_0 \rangle_B$ $\rightarrow$ $|5D_{5/2}, F = 2, m_F = 0 \rangle_B$, so that $|r_0 \rangle_B$ and $|r_1 \rangle_B$ have been shelved in the $5D_{5/2}$ hyperfine manifold after $3 \times 0.83 \: \mu$s $= 2.5 \: \mu$s.
When the transfer attempt fails, $A$ and $B$ may be left in the $|6S_{1/2}, F = 2 \rangle$ manifold, so our last step is to optically pump them via the $5P_{1/2}$ state ($<<1 \: \mu$s) out of this hyperfine manifold, in order to ensure that no stray ions couple to the cavity mode when we attempt entanglement between the next pair of communication qubits. If successful, the sequence leaves ions $A$ and $B$ shelved in a Bell state of the $5D_{5/2}$ manifold. If the sequence was unsuccessful, ion $B$ is optically pumped into either $5D_{3/2}$ or $|6S_{1/2}, F = 1 \rangle$. In either case, $A$ and $B$ are absent from $|6S_{1/2}, F = 2 \rangle$ so the cavity is cleared for another transfer attempt. Summing the amount of time taken by the two single photon transfers and the 4 single qubit rotations gives $\tau_{serial} = 2 \times 1.0 \mu$s $+4 \times 0.83 \: \mu$s $= 5.3 \: \mu$s. After all transfer attempts are completed, all communication qubits $B$ are then fluorescence-imaged on the $6S_{1/2} \rightarrow 5P_{1/2}$ transition to determine which entangling attempts were successful.

\section{\label{sec:level1} Trapped Ion Implementation: Entanglement Fidelity}

In the implementation described above the dominant expected source of error is the potential for a communication qubit to decay from the state $|e\rangle$ into either $|r_0 \rangle$ or $|r_1 \rangle$ (by emitting a photon into free space and thus destroying the Bell state), which falsely heralds a successful entanglement.
The small branching ratio from $6P_{3/2}$ to $5D_{3/2}$ of $2.9\%$ \cite{kurz2008} combined with the small branching ratio of 3/25 of the $6P_{3/2}|F = 3, m_F = 0 \rangle$ state to the particular hyperfine state $5D_{3/2}|F = 2, m_F = 0 \rangle$  strongly suppresses this backscattering. Simulating the residual excited-state populations during the transfers and incorporating the exact branching ratios, we estimate an error of about $4.5 \times 10^{-4}$ per entanglement attempt, giving an average error of $4.5 \times 10^{-4}/p = 1.1 \times 10^{-3}$ to the heralded Bell pairs. This could be further reduced if necessary through multiphoton transitions, using magnetic fields to mix hyperfine levels to take advantage of additional selection rules, or by using other species with similar level structures but even more favorable branching ratios (such as Sr$^+$, Ca$^+$).

Laser amplitude fluctuations could create mismatched heralded amplitudes $|\alpha|$ and $|\alpha'|$ on the $|00 \rangle$ and $|11 \rangle$ components of the Bell state since the success probability depends on $\Omega_0$, leading to an infidelity which scales as $\epsilon^2$ where $\epsilon = |\alpha'| - |\alpha|$. However, choosing $\Omega_0$ to have the value which optimizes the transfer efficiency for a fixed laser pulse length makes $\epsilon$ first-order insensitive to laser amplitude fluctuations. 

Similarly, $g$ varying between the two transfers could introduce errors, but slow variations are irrelevant and periodic variations on the time scale of the transfers (such as potential ion/atom trap oscillations within the cavity mode profile) can be symmetrized by synchronizing the time between pulses with the trap period. Ion motion resulting from the contributions of many collective modes is more difficult to symmetrize and is another potential source of infidelity. The severity of this variation is highly dependent on the geometry and choice of position of the communication qubit within the chain. In the geometry depicted in Fig. 1 of the main text, in which all the chains can be held in a single linear surface trap, the most variation would occur along the axial direction, where the trapping frequency is the smallest and the cavity mode profile varies the most. At a temperature of 1 mK with a center of mass axial trapping frequency of 200 kHz, we expect a corresponding thermal spread of the ion chain's position of $\sqrt{ \langle x^2 \rangle} = 32$ nm, which is appreciable compared to the 455 nm mode wavelength. We ran a simulation where we thermally populated the several lowest frequency modes \cite{james1998} and allowed the coupling strength of the communication qubits to oscillate accordingly during successive simulated transfers. While the optimal choice of ion positions for mixed species sympathetic cooling and axial/transverse gates is under investigation \cite{monroe2021}, we imagine using a few Ba$^+$ interspersed throughout the chain for efficient sympathetic cooling of many motional modes. 
Placing the Ba$^+$ ion used as the communication qubit in the center of the chain so that it has the smallest possible participation in the motion of several of the low frequency modes (and zero participation for symmetric breathing modes) minimizes the thermal motion \cite{james1998}. By additionally separating the transfers by half a period of the 200 kHz oscillation frequency (2.5 $\mu$s), we simulated an average infidelity due to thermal motion of about 1.3\%. Note that using multiple Ba$^+$ isotopes for communication and memory qubits instead of species with different masses allows maximal flexibility for choice of axial/transverse local gates and ion position \cite{monroe2021}.

While Bell state infidelity arising due to ion motion is the dominant source of error in our protocol, it is not fundamental and can be further reduced in several ways, for example, with slightly higher trapping frequencies (300 kHz center of mass frequency reduces the error to $10^{-3}$).
Similarly, pinning the communication qubits to the cavity mode antinodes by ramping up the far detuned light used to lock the cavity to form a lattice (wavelength of $2 \times 455$ nm $= 1010$ nm) along the cavity axis during the cavity transfer stage of the protocol would also reduce this effect.
Other geometric configurations (such as orienting the chains perpendicular to the cavity mode) would entirely eliminate this potential problem.

\section{\label{sec:level1}Nonlocal Gate Durations and Storage Errors}
In the main text we point out that the primary advantage of direct nonlocal gates is improved nonlocal gate fidelity due to the elimination of linearly increasing overheads of costly two-qubit gate errors.
Due to very long qubit lifetimes of neutral atoms and trapped ions, qubit storage errors are much less significant than two-qubit gate errors, but how storage errors scale with system size will in general depend not only on the degree of connectivity, but also on the degree of parallelization afforded by the architecture.

To find a general expression capturing the contributions of parallelizability and connectivity to the average nonlocal gates times and storage errors, consider how the amount of time necessary to enact nonlocal gates among all $N$ qubits in a quantum computer scales with $N$. For a system with local interactions in $D$ dimensions, each nonlocal gate requires a number of SWAP gate which scales with $N$ as $N^{1/D}$. Enacting such a gate on all $N$ qubits in the system would then require a total number of gates scaling with $N$ as $N \times N^{1/D}$, where higher dimensional connectivity clearly reduces the gate count. For a serial system, the time to enact nonlocal gates on all $N$ qubits would just scale with this gate count. With parallel operations, however, this time can be reduced by a factor of $N^*$, the number of operations that can be done in parallel, leading to an average time to nonlocally connect $N$ qubits scaling with $N$ as $N \times N^{1/D}/N^*$.

For a nearest neighbor system with $D = 1$, since the average number of SWAPs per nonlocal gate goes as $N$, even fully parallelized operations allowing $N^* \propto N$ simultaneous local gates would recover the same scaling of the average time per nonlocal gate as the serial any-to-any system, where effectively $D \rightarrow \infty$ and $N^* = 1$.
For $D = 2$ nearest neighbor systems, the average number of SWAP gates per nonlocal gate would go as $\sqrt{N}$, so scaling the number of parallelized operations as $N^* \propto \sqrt{N}$ is still necessary to equal the any-to-any serial scaling of the average gate time, although full parallelization in $2D$ would result in a $\sqrt{N}$ improvement over the any-to-any serial average nonlocal gate duration scaling.
In summary, directly implementing nonlocal gates most importantly reduces the accumulated two-qubit gate errors from increasing SWAP gate overheads, but can also help reduce average nonlocal gate durations and thereby storage errors. Nearest neighbor architectures suffer from two-qubit gate errors that scale with system size and require highly parallelized control to compete with or exceed the average nonlocal gate duration and storage error scalings in a serial any-to-any connected system.


\bibliography{ramette_biblio}